\documentclass[12pt]{iopart}
\usepackage{graphicx}
\usepackage{dcolumn}
\usepackage{bm}

\begin{document}

\title{Non-Equilibrium Josephson and Andreev Current through Interacting Quantum Dots}

\author{Marco G. Pala$^1$, Michele Governale$^2$, J\"urgen K\"onig$^2$}

\address{$^1$ IMEP-MINATEC (UMR CNRS/INPG/UJF 5130), 
38016 Grenoble, France}

\address{$^2$ Institut f\"ur Theoretische Physik III, 
Ruhr-Universit\"at Bochum, 44780 Bochum, Germany }

\eads{\mailto{pala@minatec.inpg.fr}, 
\mailto{michele@tp3.ruhr-uni-bochum.de}, 
\mailto{koenig@tp3.ruhr-uni-bochum.de}}

\begin{abstract}
We present a theory of transport through interacting quantum dots 
coupled to normal and superconducting leads in the limit of weak tunnel
coupling.
A Josephson current between two superconducting leads, carried by
first-order tunnel processes, can be established by non-equilibrium
proximity effect.
Both Andreev and Josephson current is suppressed for bias voltages below a
threshold set by the Coulomb charging energy.
A $\pi$-transition of the supercurrent can be driven by tuning gate or bias 
voltages.
\end{abstract}



\submitto{\NJP}

\maketitle

\section{Introduction}
Non-equilibrium transport through superconducting systems attracted much 
interest since the demonstration of a 
Superconductor-Normal-Superconductor (SNS) transistor \cite{baselmans99}. 
In such a device, 
supercurrent suppression and its sign reversal ($\pi$-transition)
are achieved by driving the
quasi-particle distribution out of equilibrium
by means of applied voltages \cite{volkov95, wilhelm98, yip98,giazotto04}. 
Another interesting issue in mesoscopic physics is transport through quantum 
dots attached to superconducting leads.
For DC transport through quantum dots coupled to a normal and a 
superconducting lead, subgap transport is due to Andreev 
reflection \cite{fazio98, kang98, schwab99, clerk00, lambert00, cuevas01}.
Also transport between two superconductors through a quantum dot has been 
studied extensively.
The limit of a non-interacting dot has been investigated in \cite{beenakker92}.  
Several authors considered the regime of weak tunnel coupling where the electrons forming a Cooper pair 
tunnel one by one via virtual states \cite{glazman89,spivak91,rozhkov01}.  The Kondo regime was 
also addressed \cite{glazman89,ambegaokar00,avishai03,sellier05,lopez07}.  
Multiple Andreev reflection through localized levels was investigated in ~\cite{cuevas97,johansson99}. 
Numerical approaches based on the non-crossing approximation \cite{ando95}, 
the numerical renormalization group \cite{choi04} and Monte 
Carlo \cite{siano04} 
have also been used. The authors of~\cite{vecino03} compare different approximation schemes, such as mean field and second-order perturbation in the Coulomb interaction. 
In double-dot systems the Josephson current has been shown to depend on the spin state 
of the double dot \cite{choi00}.
Experimentally, the supercurrent through a quantum dot has been measured through dots realized in 
carbon nanotubes \cite{nanotubes} and in indium arsenide nanowires \cite{nanowires}.

In this Letter we study the transport properties of a system composed of an 
interacting single-level quantum dot between two equilibrium superconductors 
where a third, normal lead is used to drive the dot out of equilibrium. 
A Josephson coupling in SNS heterostructures can be mediated by 
proximity-induced superconducting correlations in the normal region.
In case of a single-level quantum dot, superconducting correlations are 
indicated by the correlator $\langle d_\downarrow (0) d_\uparrow (t) \rangle$,
where $d_\sigma$ is the annihilation operator of the dot level with spin 
$\sigma$.
To obtain a large pair amplitude, i.e. the equal-time correlator 
$\langle d_\downarrow d_\uparrow \rangle$, at least two conditions need to be 
fulfilled: (i) the states of an empty and a doubly-occupied dot should be 
nearly energetically degenerate and (ii) the overall probability of occupying the dot
with an even number of electrons should be finite.
For a non-interacting quantum dot, i.e. vanishing charging energy $U$ for 
double occupancy, this can be achieved by tuning the level position $\epsilon$
in resonance with the Fermi energy of the leads, $\epsilon=0$ 
\cite{beenakker92}.
In this case, the Josephson current can be viewed as transfers of Cooper pairs
between dot and leads and the expression of the current starts in first order 
in the tunnel coupling strength $\Gamma$.

The presence of a large charging energy $U \gg k_BT, \Gamma$ destroys this 
mechanism since the degeneracy condition $2\epsilon + U \approx 0$ is 
incompatible with a finite equilibrium probability to occupy the dot with 
an even number of electrons.
Nevertheless, a Josephson current can be established by higher-order 
tunnelling processes (see, for example, \cite{glazman89,spivak91,rozhkov01}), 
associated with a finite superconducting correlator 
$\langle d_\downarrow (0) d_\uparrow (t) \rangle$ at different times.
The amplitude of the Josephson coupling is, however, reduced by a factor 
$\Gamma/\Delta$, i.e., the current starts only in second order in $\Gamma$,
and the virtual generation of quasiparticles in the leads suppresses the 
Josephson current for large superconducting gaps $\Delta$. 
In particular, it vanishes for $\Delta \rightarrow \infty$.

The main purpose of the present paper is to propose a new mechanism that 
circumvents the above-stated hindrance  
to achieve a finite pair amplitude in 
an interacting quantum dot, and, thus, restores a Josephson current
carried by first-order tunnel processes that survives in the limit
$\Delta \rightarrow \infty$.
For this aim, we attach a third, normal, lead to the dot that drives the latter
out of equilibrium by applying a bias voltage, so that condition of occupying 
the dot with an even number of electrons is fulfilled even for
$2\epsilon+U \approx 0$.

We relate the current flowing into the superconductors to the nonequilibrium 
Green's functions of the dot.
In the limit of a large superconducting gap, $\Delta \rightarrow \infty$,
the current is only related to the pair amplitude.
The latter is calculated by means of a kinetic equation derived from a 
systematic perturbation expansion within real-time diagrammatic technique 
that is suitable for dealing with both strong Coulomb interaction and
nonequilibrium at the same time.

\section{Model}
\begin{figure}[h]
\begin{center}
\includegraphics{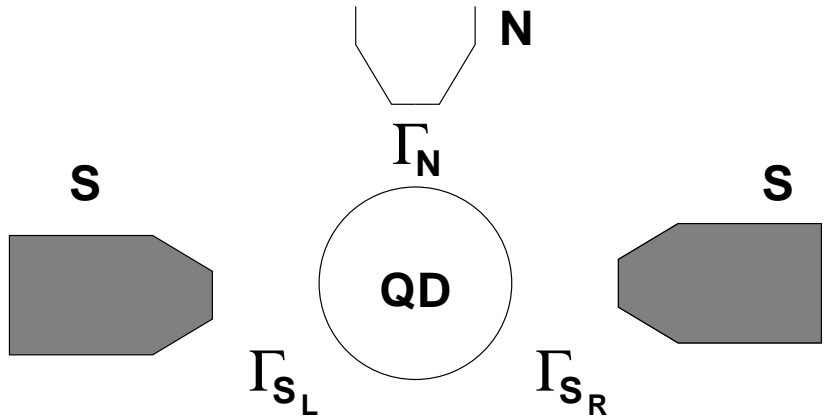}
\caption{
Setup: a single-level quantum dot is connected by tunnel junctions
 to one normal and two superconducting leads with tunnelling rates 
$\Gamma_{\rm N}$ and $\Gamma_{{\rm S}_{{\rm L,R}}}$, respectively.\label{figure1}}
\end{center}
\end{figure}

We consider a single-level quantum dot 
connected to two superconducting and one normal lead
via tunnel junctions, see figure~\ref{figure1}.
The total Hamiltonian is given by $H=H_{{\rm D}}+\sum_{\eta= {\rm N}, {\rm S}_{L},{\rm S}_{R}}( H_\eta +H_{{\rm tunn},\eta})$.
The quantum dot is described by the Anderson model 
$H_{{\rm D}}=\sum_{\sigma} \epsilon d_{\sigma}^\dagger d_{\sigma}+
U n_{\uparrow} n_{\downarrow}$, where 
$n_{\sigma}=d_{\sigma}^\dagger d_{\sigma}$ is the number operator for spin 
$\sigma=\uparrow,\downarrow$, $\epsilon$ is the energy level, 
and $U$ is the charging energy for double occupation.
The leads, labeled by $\eta= {\rm N}, {\rm S_L}, {\rm S_R}$, are modeled by 
$H_{\eta}=\sum_{k \sigma} 
\epsilon_{k}c_{\eta k \sigma}^\dagger c_{\eta k \sigma} - \sum_k \left( 
\Delta_\eta c_{\eta k \uparrow}^\dagger c_{\eta -k \downarrow}^\dagger 
+{\rm H.c.} \right)$, where $\Delta_\eta$ is the superconducting order 
parameter ($\Delta_{\rm N}=0$).
The tunnelling Hamiltonians are 
$H_{{\rm tunn},\eta}= V_{\eta} \sum_{k \sigma} \left( c_{\eta k \sigma}^\dagger 
d_\sigma +{\rm H.c.} \right)$.
Here, $V_{\eta}$ are the spin- and wavevector-independent tunnel matrix 
elements, and $c_{\eta k \sigma} (c_{\eta k \sigma}^\dagger)$ and 
$d_\sigma (d_\sigma^\dagger)$ represent 
the annihilation (creation) operators for the leads and dot, respectively.
The tunnel-coupling strengths are characterized by
$\Gamma_{\eta}=2 \pi |V_{\eta}|^2 
\sum_k  \delta (\omega - \epsilon_k)$.

\section{Current formula}

We start with deriving a general formula for the charge current in
lead $\eta$ by using the approach of Ref.~\cite{meir-wingreen} 
generalized to superconducting leads.
Similar formulae that relate the charge current to the Green's function of the 
dot in the presence of superconducting leads have been derived in previous 
works, in particular for equilibrium situations, see e.g. 
Refs.~\cite{ambegaokar00,ando95}.
The formula derived below is quite general, as it allows for arbitrary bias
and gate voltages, temperatures, and superconducting order parameters for a 
quantum dot coupled to an arbitrary number of normal and superconducting leads.
For this, it is convenient to use the operators $\psi_{\eta k}=( 
c_{\eta k \uparrow}, c_{\eta -k \downarrow}^{\dagger})^{{\rm T}}$
and $\phi=( d_{\uparrow}, d^{\dagger}_{\downarrow})^{{\rm T}}$
in Nambu formalism.
The current from lead $\eta$ is expressed as
$J_{\eta} = e \left\langle d N_\eta / dt \right\rangle
= i(e/\hbar) \langle [H,N_\eta]\rangle
= i(e/\hbar) \langle [H_{{\rm tunn},\eta},N_\eta]\rangle$
\footnote{Note that $[H_\eta,N_\eta] \neq 0$ but $\langle [H_\eta,N_\eta] \rangle = 0$
for $\eta={\rm S}_{\rm L,R}$.},
with $N_\eta=\sum_k \psi^{\dagger}_{\eta k}\tau_3 \psi_{\eta k}$,
where $\tau_1, \tau_2, \tau_3$ indicate the Pauli matrices in Nambu space
and $e>0$ the electron charge.
Evaluating the commutator leads to
\begin{equation}
  J_{\eta} = - \frac{2 e}{ \hbar}\sum_k\int \frac{d \omega}{2\pi}{\rm Re}
  \left\{{\rm Tr} \left[\tau_3 \mathbf{V}_{\eta} 
    \mathbf{G^{<}}_{{\rm D}, \eta  k}(\omega)\right]\right\} , 
  \label{current1}
\end{equation}
with $\mathbf{V}_{\eta}={\rm Diag}(V_\eta, -V^{*}_\eta)$
and the lead--dot lesser Green's functions
$\left( \mathbf{G^{<}}_{{\rm D}, \eta k}(\omega)\right)_{m, n}$ that are
the Fourier transforms of $i\langle\psi^{\dagger}_{\eta k n}(0)\phi_m(t)
\rangle$. In the following, we assume the tunnelling matrix elements $V_\eta$ 
to be real 
(any phase of $V_\eta$ can be gauged away by substituting
$\Delta_\eta \rightarrow \Delta_\eta \exp (-2i\arg V_\eta)$).
The Green's function $\mathbf{G^{<}}_{{\rm D}, \eta k}$ is related to the full dot Green's functions and 
the lead Green's functions by a Dyson equation in Keldysh formalism: $\mathbf{G^{<}}_{{\rm D},\eta k}(\omega)=\mathbf{G}^{{\rm R}}(\omega)
  \mathbf{V}^{\dagger}_{\eta}\mathbf{g}^{<}_{\eta k}(\omega)+ 
  \mathbf{G}^{<}(\omega)
  \mathbf{V}^{\dagger}_{\eta}\mathbf{g}^{{\rm A}}_{\eta k}(\omega)$, 
where  $\mathbf{G}^{{\rm R}(<)}(\omega)$ is the retarded (lesser) 
dot Green's function, and  
and $\mathbf{g}^{{\rm A}(<)}_{\eta k}(\omega)$ 
the lead advanced (lesser) Green's function. 
Using this relation and assuming energy-independent tunnel rates $\Gamma_\eta$, 
we obtain for the current $J_\eta = J_{1\eta} + J_{2\eta}$ with 
\begin{eqnarray}
\label{j1}
  J_{1\eta} &=& \frac{e}{\hbar} \int \frac{d \omega}{2\pi}
  \Gamma_{\eta} D_\eta(\omega)
	{\rm Im} \left\{ {\rm Tr} \left[ \tau_3 
	  \left[\mathbf{1} - 
   \frac{\mathbf{\Delta}_\eta}{\omega-\mu_{\eta}} \right]
    \right. \right. \nonumber \\ && \left. \left.
    \left(2 \mathbf{G}^{{\rm R}}(\omega) f_\eta(\omega)+ 
    \mathbf{G}^{<}(\omega)\right) \right]\right\} ,\\
  J_{2\eta} &=& \frac{e}{\hbar} \int \frac{d \omega}{2 \pi} 
  \Gamma_{\eta} S_\eta(\omega)
  {\rm Re}\left\{{\rm Tr}\left[ \tau_3 \frac{\mathbf{\Delta}_\eta}{|\Delta_\eta |}
  \mathbf{G}^{<}(\omega)\right]\right\} ,
  \label{current2}
\end{eqnarray} 
where 
$\mathbf{\Delta}_\eta = \left( \begin{array}{cc} 0 & \Delta_\eta \\ 
\Delta_\eta^* & 0 \end{array} \right)$,
and $f_\eta(\omega)=[1+\exp(\omega-\mu_\eta)/ (k_{{\rm B}} T)]^{-1}$ is the Fermi function, with $T$ being the temperature and $k_{{\rm B}}$ the Boltzmann constant. 
 The dot Green's functions $\left( \mathbf{G^{<}}_{{\rm D}}(\omega)\right)_{m, n}$ and $\left( \mathbf{G^{{\rm R}}}_{{\rm D}}(\omega)\right)_{m, n}$ are defined as 
the Fourier transforms of $i\langle\phi^{\dagger}_{n}(0)\phi_m(t)
\rangle$ and $-i \theta(t) \langle \{ \phi_{m}(t),\phi^{\dagger}_{n}(0) \}\rangle$, respectively.
The two weighting functions $D_\eta (\omega)$ and $S_\eta (\omega)$ are given by 
\begin{eqnarray}
\nonumber
D_\eta(\omega)&=&\frac{|\omega-\mu_\eta|}{\sqrt{(\omega-\mu_\eta)^2 -|\Delta_\eta|^2}}\theta(|\omega-\mu_\eta |-|\Delta_\eta|)\\  
\nonumber
S_\eta(\omega)&=&\frac{|\Delta_\eta |}
	{\sqrt{|\Delta_\eta|^2-(\omega-\mu_\eta)^2}}\theta(|\Delta_\eta|-|\omega-\mu_\eta |).
\end{eqnarray} 
The terms $J_{1\eta}$ and $J_{2\eta}$ involve excitation energies $\omega$ 
above and below the superconducting gap, respectively.
For $\eta={\rm N}$, only the part of $J_{1\eta}$ that involves normal 
(diagonal) components of the Green's functions contributes, and the current 
reduces to the result presented in \cite{meir-wingreen}.
For superconducting leads, this part describes quasiparticle transport that
is independent of the superconducting phase difference.
The other part of $J_{1\eta}$ involves anomalous (off-diagonal) components of 
the Green's functions and is, in general, phase dependent.
The contribution to the Josephson current stemming from this term is the 
dominant one in the regime considered in \cite{glazman89,spivak91,rozhkov01}. 
The excitation energies above the gap are only accessible either for transport 
voltages exceeding the gap or by including higher-order tunnelling, involving 
virtual states with quasiparticles in the leads, and, therefore, 
$J_{1\eta}$ vanishes for large $|\Delta_\eta|$.
In this case $J_{2\eta}$, that involves only anomalous Green's functions
with excitation energies below the gap, dominates transport.
It is, in general, phase dependent, and describes both Josephson as well as 
Andreev tunnelling.

In the following we consider the limit $|\Delta_\eta| \rightarrow \infty$, 
where the current simplifies to
\begin{equation}
\label{jeta}
  J_\eta = \frac{2 e}{\hbar} \Gamma_\eta |\langle d_\downarrow d_\uparrow \rangle| 
\sin (\Psi - \Phi_\eta) \, ,
\end{equation}
with $\Phi_\eta$ being the phase of $\Delta_\eta$  and 
$\langle d_\downarrow d_\uparrow \rangle=| \langle d_\downarrow d_\uparrow \rangle| 
\exp(i \Psi)$
the pair amplitude of the dot that has to be determined in the presence of Coulomb interaction, coupling to all (normal and superconducting) leads and in non-equilibrium due to finite bias voltage.

We now consider a symmetric three-terminal setup with
$\Gamma_{{\rm S}_{{\rm L}}}=\Gamma_{{\rm S}_{{\rm R}}}=\Gamma_{{\rm S}}$, $\Delta_{{\rm S}_{{\rm L}}} = |\Delta| \exp ( i \Phi/2)$ and
$\Delta_{{\rm S}_{{\rm R}}} = |\Delta| \exp ( -i \Phi/2)$, and $\mu_{{\rm S}_{{\rm L}}}=\mu_{{\rm S}_{{\rm R}}}=0$. 
The quantities of interest are the the current that flows between the two superconductors  (Josephson current) $J_{\rm jos} = (J_{{\rm S}_{{\rm L}}} - J_{{\rm S}_{{\rm R}}})/2$ and the 
current in the normal lead 
(Andreev current) $J_{\rm and} = J_{{\rm N}} = - (J_{{\rm S}_{{\rm L}}} + J_{{\rm S}_{{\rm R}}}) $.

Furthermore, we focus on the limit of weak tunnel coupling, 
$\Gamma_{\rm S} < k_{\rm B}T$.
In this regime, an Josephson current through the dot in equilibrium 
would be suppressed even in the absence of Coulomb interaction, $U=0$, 
since the influence of the superconductors on the quantum-dot spectrum 
could not be resolved for the resonance condition $\epsilon \approx 0$.
This can, e.g., be seen in the exactly-solvable limit of
$U=0$ together with  $\Gamma_{\rm N}=0$, where the Josephson current is
$J_{{\rm jos}} = (e/2\hbar) \Gamma_{{\rm S}}^2 \sin(\Phi)
\left[f(-\epsilon_{{\rm A}}(\Phi)) -f(\epsilon_{{\rm A}}(\Phi))\right]
/\epsilon_{{\rm A}}(\Phi)$
with $\epsilon_{{\rm A}}(\Phi)=\sqrt{\epsilon^2+\Gamma_{{\rm S}}^2 
\cos^2 (\Phi/2)}$.
This provides an additional motivation to look for a {\it non-equilibrium} 
mechanism to proximize the quantum dot.

\section{Kinetic equations for quantum-dot degrees of freedom} 
The Hilbert space of the dot is four dimensional: the dot can be 
empty, singly occupied with spin up or down, or doubly occupied, denoted by
$|\chi\rangle\in \{| 0 \rangle$, $| \uparrow \rangle$, $| \downarrow \rangle, | D \rangle\equiv d^{\dagger}_{\uparrow} d^{\dagger}_{\downarrow}|0\rangle \}$, with energies $E_0$, $E_\uparrow=E_\downarrow$, $E_{D}$. 
For convenience we define the detuning as 
$\delta=E_{D}-E_0=2 \epsilon+U$. 
The dot dynamics is fully described by its reduced density matrix
$\rho_{\rm D}$, with matrix elements $P_{\chi_2}^{\chi_1} \equiv (\rho_{\rm D})_{\chi_2 \chi_1}$. 
The dot pair amplitude $\langle d_\downarrow d_\uparrow \rangle$ is given by 
the off-diagonal matrix element $P_D^0$. 
The time evolution of the reduced density matrix is described by the 
kinetic equations
\begin{equation}
  \frac{d}{d t} P^{\chi_1}_{\chi_2}(t) 
  +\frac{i}{\hbar} (E_{\chi_1}-E_{\chi_2}) P^{\chi_1}_{\chi_2}(t)
  = \sum_{\chi_1', \chi_2'} \int_{t_0}^{t} dt' 
  W^{\chi_1 \chi_1'}_{\chi_2 \chi_2'} (t,t') P^{\chi_1'}_{\chi_2'} (t')
  \label{te}  .
\end{equation}
We define the generalized transition rates by 
$W^{\chi_1 \chi_1'}_{\chi_2 \chi_2'} 
\equiv \int_{-\infty}^t dt' W^{\chi_1 \chi_1'}_{\chi_2 \chi_2'} (t,t')$, which
are the only quantities to be evaluated in the stationary limit.
Together with the normalization condition
$\sum_\chi P_\chi=1$, (\ref{te}) determines the matrix elements of 
$\rho_{\rm D}$.
Furthermore, in (\ref{te}) we retain only linear terms in the tunnel strengths 
$\Gamma_\eta$ and the detuning $\delta$.
Hence, we calculate the rates $W^{\chi_1 \chi_1'}_{\chi_2 \chi_2'}$ 
to the lowest (first) order in $\Gamma_\eta$ for $\delta=0$.
This is justified in the transport regime 
$\Gamma_{\rm S}, \Gamma_{\rm N}, \delta < k_{{\rm B}}T$. 

The rates are evaluated by means of a real-time diagrammatic technique 
\cite{koenig96}, that we generalize to include superconducting leads.
This technique provides a convenient tool to perform a systematic 
perturbation expansion of the transport properties in powers of the 
tunnel-coupling strength.
In the following, we concentrate on transport processes to first order in 
tunnelling (a generalization to higher orders is straightforward).
This includes the transfer of charges through the tunnelling barriers
as well as energy-renormalization terms that give rise to nontrivial dynamics
of the quantum-dot degrees of freedom.

We find for the (first-order) diagonal rates 
$W_{\chi_1 \chi_2}\equiv W_{\chi_1 \chi_2}^{\chi_1 \chi_2}$ the expressions
$W_{\sigma 0}=\Gamma_{\rm N} f_{\rm N}(-U/2); 
W_{0 \sigma}=\Gamma_{\rm N} [1-f_{\rm N}(-U/2)];
W_{D \sigma}=\Gamma_{\rm N} f_{\rm N} (U/2);
W_{\sigma D}=\Gamma_{\rm N} [1-f_{\rm N} (U/2)]$.
The N lead also contributes to the rates   
$W_{0 0}^{D D}=(W_{D D}^{0 0})^*=- \Gamma_{\rm N} [1+f_{\rm N} (-U/2)-f_{\rm N}(U/2) +i B]$ 
where
$  B=\frac{1}{\pi} {\rm Re} \left[ \psi \left( \frac{1}{2}
    +i\frac{U/2-\mu_{{\rm N}}}{2\pi k_{\rm B}T} \right) - \psi \left(\frac{1}{2}
    +i\frac{-U/2-\mu_{{\rm N}}}{2\pi k_{\rm B}T} \right) \right] ,
$
with $\mu_{{\rm N}}$ being the chemical potential of the normal lead 
and $\psi(z)$ the Digamma function.
Notice that $B$ 
vanishes when $\mu_{{\rm N}}=0$ or $U=0$.
The superconducting leads do not enter here due to the gap in the 
quasi-particle density of states.
These leads, though, contribute to the off-diagonal rates
$W_{0 D}^{0 0} = W_{0 D}^{D D} = \left(W_{0 0}^{0 D}\right)^* =
\left( W_{D D}^{0 D} \right)^* = - W_{0 0}^{D 0} = - W_{D D}^{D 0} =
- \left( W_{D 0}^{0 0} \right)^* = - \left( W_{D 0}^{D D} \right)^* =
-i \Gamma_{{\rm S}} \cos( \Phi/2)$.

For an intuitive representation of the system dynamics
we define, in analogy to \cite{braun04}, a dot isospin by
\begin{equation}
I_x=\frac{P_0^D+P_D^0}{2}; \; I_y=i\frac{P_0^D-P_D^0}{2}; \; 
I_z=\frac{P_D-P_0}{2}.
\end{equation}
From (\ref{te}), we find that in the stationary limit  
the isospin dynamics can be separated into three parts,
$ 0= d {\bf I}/dt=\left( d {\bf I}/dt \right)_{\rm acc}+\left( d {\bf I}/dt 
\right)_{\rm rel}+ \left( d {\bf I}/dt \right)_{\rm rot}$,
with
\begin{eqnarray}
\label{iaccum}
  \hbar \left( \frac{d {\bf I}}{dt} \right)_{\rm acc}&=& 
  -\frac{\Gamma_{\rm N}}{2}[1-f_{\rm N}(-U/2) - f_{\rm N}(U/2)]
  \hat{{\bf e}}_z\\
\label{irel}
  \hbar \left( \frac{d {\bf I}}{dt} \right)_{\rm rel}&=& -\Gamma_{\rm N}
       [1+f_{\rm N}(-U/2) - f_{\rm N}(U/2) ] {\bf I}\\
  \hbar \left( \frac{d {\bf I}}{dt} \right)_{\rm rot}&=&
       {\bf I} \times {\bf B}_{\rm eff}
       \label{irot} 
\end{eqnarray}
where  $ \hat{{\bf e}}_z$ is the $z$-direction and
${\bf B}_{\rm eff}=
\left\{2 \Gamma_{\rm S} \cos (\Phi/2), 0, -\Gamma_{\rm N}B-2 \epsilon-U \right\}$
is an effective magnetic field in the isospin space. 
The accumulation term (\ref{iaccum}) builds up a finite isospin,
while the relaxation term (\ref{irel}) decreases it.
Finally, (\ref{irot}) describes a rotation of the isospin direction.

\section{Non-equilibrium Josephson current}
In the isospin language the current in  the superconducting leads is
\begin{equation}
J_{{\rm S}_{{\rm L,R}}} = \frac{2e}{\hbar} 
  \Gamma_{\rm S}\left[ I_y \cos(\Phi/2)\pm I_x \sin(\Phi/2)\right],
\end{equation}
where the upper(lower) sign refers to the left(right) superconducting lead. 
The $I_y$ component contributes to the Andreev current, while $I_x$ is 
responsible for the Josephson current.
To obtain subgap transport, we first need to build up a finite isospin component along 
the $z$-direction, i.e. we need a population imbalance between the empty and doubly occupied dot 
[(this is generated by the accumulation term in (\ref{iaccum})]; 
second, we need a finite ${\bf B}_{\rm eff}$ which rotates the isospin so that it acquires an 
inplane component. In order to have a finite Josephson current ($I_x\ne 0$), we need the $z$-component, $-\Gamma_{\rm N}B-2 \epsilon-U$, of the effective magnetic field  
producing the rotation to be non zero.

\begin{figure}
\begin{center}
\includegraphics[width=0.8\columnwidth]{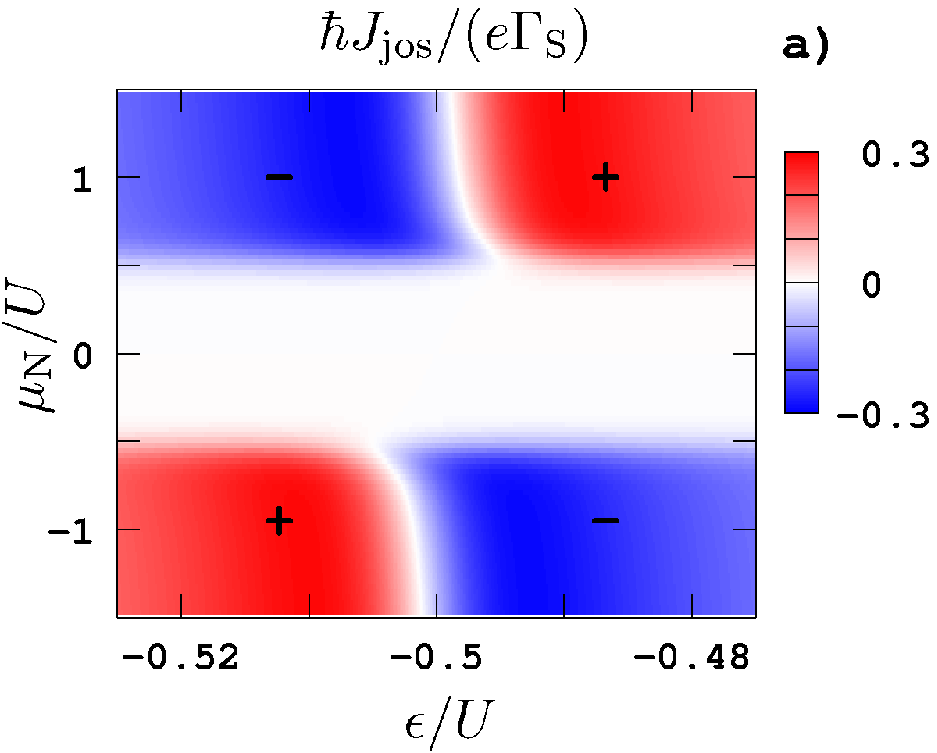}\\ \includegraphics[width=0.8\columnwidth]{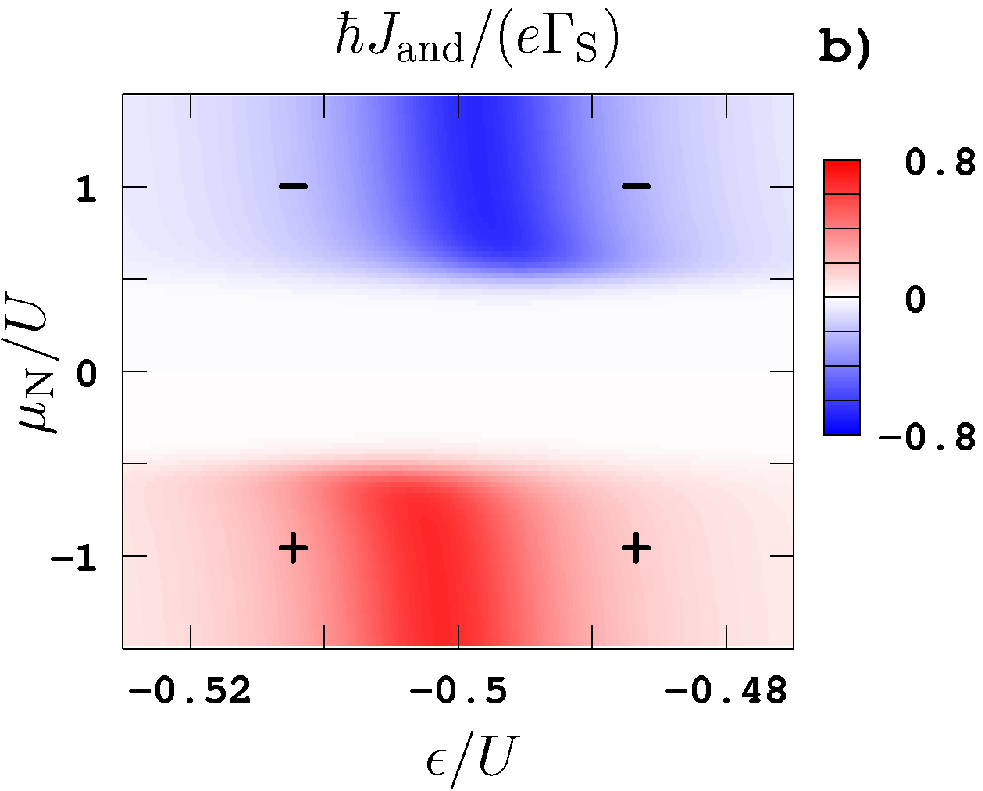}
\caption{
  Density plot of the a) Josephson and b) Andreev current, for fixed
  superconducting-phase difference $\Phi=\pi/2$, as a function of the dot-level
  position $\epsilon$ and of the chemical potential of the normal lead
  $\mu_{{\rm N}}$.
  The symbols $\pm$ refer to the sign of the current.
  The other parameters are $\Gamma_{{\rm S}}=\Gamma_{{\rm N}}=0.01 U$, and
  $k_{{\rm B}}T=0.05 U$.
\label{figure2}}
\end{center}
\end{figure}

\begin{figure}
\begin{center}
\includegraphics[width=0.8\columnwidth]{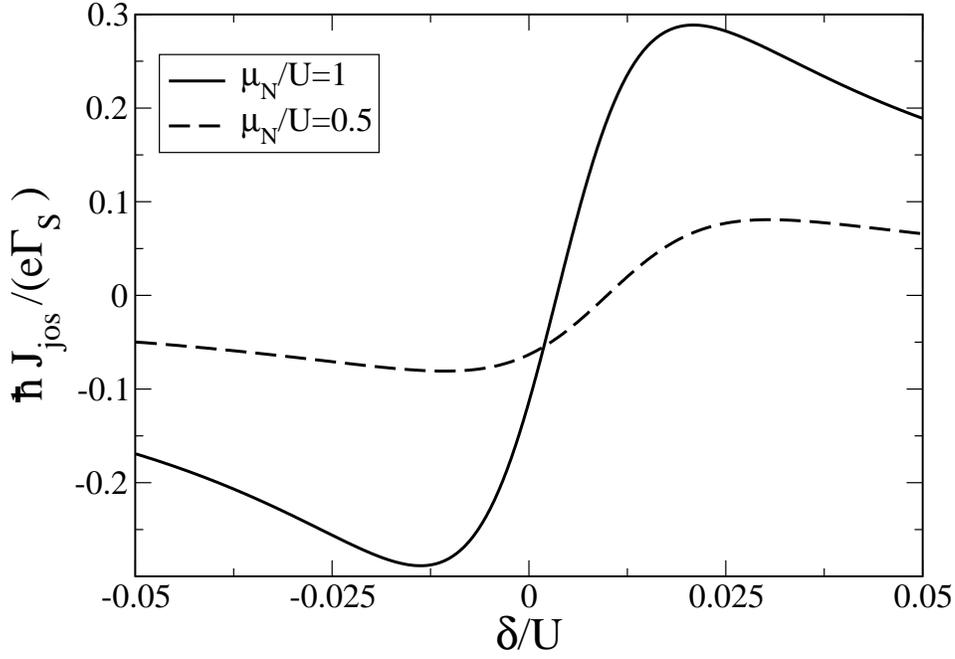}
\caption{Josephson current, for fixed superconducting-phase difference $\Phi=\pi/2$, 
as a function of the detuning $\delta=E_{D} -E_0=2\epsilon+U$ for different values of the chemical potential. 
The other parameters are $\Gamma_{\rm S}=\Gamma_{\rm N}=0.01 U$ and  
$k_{{\rm B}}T=0.05 U$.
\label{figure3}}
\end{center}
\end{figure}

The  Josephson current and the Andreev current read
\begin{eqnarray}
  \nonumber
  J_{{\rm jos}}& =&
  - \frac{e \Gamma_{\rm S} }{\hbar} \frac{[2\epsilon+U+\Gamma_{\rm N}B] 
    \Gamma_{\rm S} \sin(\Phi)}
  {|{\bf B}_{\rm eff}|^2+\Gamma_{{\rm N}}^2 
    [1+f_{\rm N}(-U/2) - f_{\rm N}(U/2)]^2
  } \\
  & &\times \frac{1-f_{\rm N}(-U/2)- f_{\rm N}(U/2)}
  {1+f_{\rm N}(-U/2)- f_{\rm N}(U/2)} \\
  \label{jc}
  \nonumber
  J_{{\rm and}}& =&
  \frac{e \Gamma_{{\rm S}} }{\hbar} \frac{2 \Gamma_{{\rm N}} \Gamma_{{\rm S}} 
    [1+\cos(\Phi)] }
  { |{\bf B}_{\rm eff}|^2+\Gamma_{{\rm N}}^2 
    [1+f_{\rm N}(-U/2) - f_{\rm N}(U/2)]^2
  }
  \\
  & & \times [1-f_{\rm N}(-U/2)- f_{\rm N}(U/2)].
  \label{ac}
\end{eqnarray}
These results take into account only first-order tunnel processes, i.e. 
the rates $W^{\chi_1 \chi_1'}_{\chi_2 \chi_2'}$ are computed to first order in $\Gamma_\eta$. 
The factor $[1-f_{\rm N}(-U/2)- f_{\rm N}(U/2)]$ ensures that no finite dot-pair 
amplitude can be established if the chemical potential of the normal lead, $
\mu_{{\rm N}}$, is inside the interval $[-U/2, U/2 ]$ 
by at least $k_{{\rm B}} T$. In this situation both the Josephson and the Andreev currents 
vanish. On the other hand, this factor takes the value $-1$ if $\mu_{{\rm N}}> U/2$ 
and the value $+1$ if   $\mu_{{\rm N}}< -U/2$. 
Hence, the sign of the Josephson current can be reversed by the applied voltage (voltage driven $\pi$-transition).  
The considerations above establish the importance of the non-equilibrium voltage 
to induce and control proximity effect in the interacting quantum dot. 
In figure \ref{figure2} we show in a density plot (a) $J_{{\rm jos}}$ and
(b) $J_{{\rm and}}$ for $\Phi=\pi/2$ as a function of the voltage $\mu_{{\rm N}}$ and the level position $\epsilon$. 
Both the control of proximity effect by the chemical potential 
$\mu_{{\rm N}}$ and the voltage driven $\pi$-transition are clearly visible. 
If the detuning is too large, 
$|\delta+\Gamma_{{\rm N}} B|> \sqrt{\Gamma_{{\rm N}}^2 +4 \Gamma_{{\rm S}}^2 \cos^2(\Phi/2)}$, 
it becomes difficult to build a superposition of the states $|0\rangle$ and $|D\rangle$, which is 
necessary to establish proximity.
As a consequence, the Josephson and the Andreev current are algebraically
suppressed by $\delta^{-1}$ and $\delta^{-2}$, respectively.
Figure~\ref{figure3} shows the Josephson current as a function of $\delta=2\epsilon+U$.  
The fact that the Josephson current is non zero 
for $\delta=0$ is due to the term $\Gamma_{{\rm N}} B$, i.e. of the interaction induced contribution to the 
$z$-component of the effective field $\mathbf{B}_{{\rm eff}}$ acting on the isospin. 
The term  $|B|$ has a maximum at $\mu_{{\rm N}}=U/2$, 
which causes this effect to be more pronounced at the onset of transport.    
The fact that the value of the Josephson current varies on a scale smaller than
temperature indicates its nonequilibrium nature.

\begin{figure}
\begin{center}
\includegraphics[width=0.8\columnwidth]{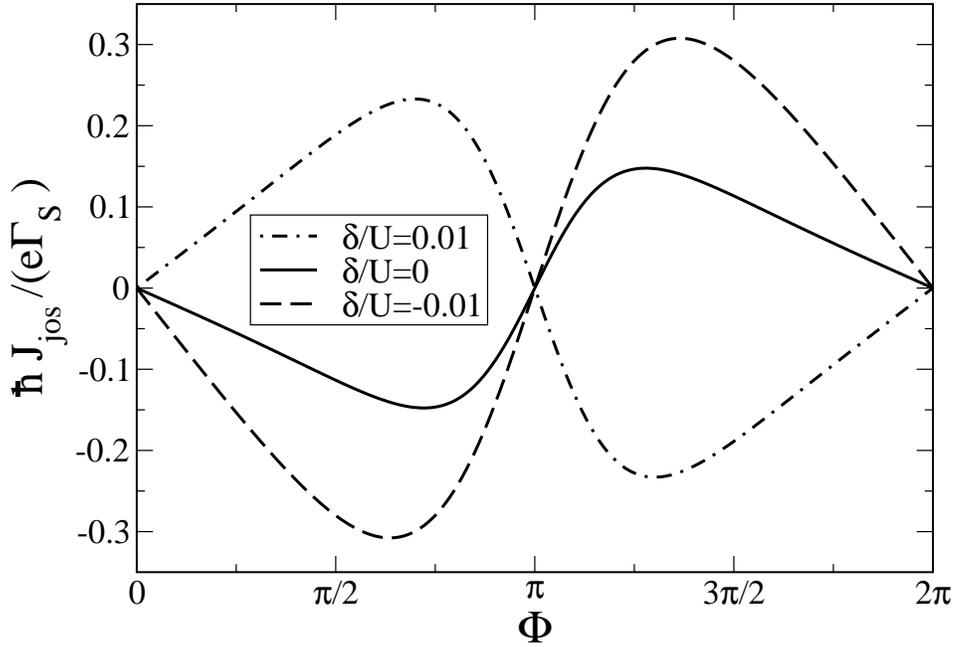}
\caption{Josephson current as a function of the superconducting-phase 
difference $\Phi$ for different values of the detuning. 
The other parameters are $\Gamma_{\rm S}=\Gamma_{\rm N}=0.01 U$, $\mu_{{\rm N}}=U$, and  
$k_{{\rm B}}T=0.05 U$.
\label{figure4}}
\end{center}
\end{figure}

A $\pi$-transition of the Josephson current can also be achieved by changing 
the sign of $\delta+\Gamma_{\rm N}B$, as shown in figure~\ref{figure4} where 
$J_{{\rm jos}}$ is plotted as a function of the phase difference $\Phi$ for 
different values of the level position. 
Notice that the current for $\delta=0$ ($\epsilon=-U/2$) 
is different from zero only due to the presence of the term $\Gamma_{{\rm N}} B$ acting on the isospin.  

\section{Conclusion}
In conclusion, we have studied non-equilibrium proximity effect in an 
interacting single-level quantum dot weakly coupled to two superconducting 
and one normal lead.
We propose a new mechanism for a Josephson coupling between the leads that 
is qualitatively different from earlier proposals based on higher-order 
tunnelling processes via virtual states. 
Our proposal relies on generating a finite {\it non-equilibrium} pair amplitude
on the dot by applying a bias voltage between normal and superconducting leads.
The charging energy of the quantum dot defines a {\it threshold bias voltage} 
above which the non-equilibrium proximity effect allows for a Josephson 
current carried by {\it first-order} tunnelling processes, that is 
{\it not suppressed} in the limit of a large superconducting gap.
Both the {\it magnitude} and the {\it sign} of the Josephson current are 
sensitive to the energy difference between empty and doubly-occupied dot. 
A {\it $\pi$-transition} can be driven by either bias or gate voltage.
In addition to defining a threshold bias voltage, the charging energy
induces many-body correlations that affect the dot's pair amplitude,
visible in a {\it bias-voltage-dependent shift} of the $\pi$-transition as 
a function of the gate voltage.
  
\ack{We would like to thank W. Belzig, R. Fazio, A. Shnirman, and A. Volkov for useful discussions. 
M.G. and J.K. acknowledge the hospitality of Massey University, 
Palmerston North, and of the CAS Oslo, respectively.}

\section*{References}

\end{document}